\documentclass[twocolumn,aps,superscriptaddress,pre]{revtex4}

\usepackage{amsmath,amssymb,graphicx}
\usepackage{algorithmic}
\usepackage{enumerate}
\usepackage{times}
\usepackage{color}
\usepackage{soul}
\definecolor{yblue}{rgb}{0.06, 0.3, 0.57}
\usepackage[pdftex]{hyperref}
\hypersetup{colorlinks=true,linkcolor=blue,citecolor=blue,urlcolor=blue}

\newcommand{\mean}[1]{\langle{#1}\rangle}

\begin{document}

\title{Three-dimensional universality class of Ising model with power-law-correlated critical disorder}

\author{Wenlong Wang}
\email{wenlongcmp@gmail.com}
\affiliation{Department of Physics, Royal Institute of Technology, Stockholm, SE-106 91, Sweden}

\author{Hannes Meier}
\affiliation{Department of Physics, Royal Institute of Technology, Stockholm, SE-106 91, Sweden}

\author{Jack Lidmar}
\affiliation{Department of Physics, Royal Institute of Technology, Stockholm, SE-106 91, Sweden}

\author{Mats Wallin}
\affiliation{Department of Physics, Royal Institute of Technology, Stockholm, SE-106 91, Sweden}

\begin{abstract}
We use large-scale Monte Carlo simulations to test the Weinrib-Halperin criterion that predicts new universality classes in the presence of sufficiently slowly decaying power-law-correlated quenched disorder. 
While new universality classes are reasonably well established, the predicted exponents are controversial. We propose a 
method of growing such correlated disorder using the three-dimensional Ising model as benchmark systems both for generating disorder and studying the resulting phase transition. Critical equilibrium configurations of a disorder-free system are used to define the two-value distributed random bonds
with a small power-law exponent given by the pure Ising exponent. 
Finite-size scaling analysis shows a new universality class with a single phase transition, but the critical exponents 
$\nu_d=1.13(5), \eta_d=0.48(3)$
differ significantly from theoretical predictions. 
We find that depending on details of the disorder generation, disorder-averaged quantities can develop peaks at two temperatures for finite sizes.
Finally, a layer model with the two values of bonds spatially separated to halves of the system genuinely has multiple phase transitions and 
thermodynamic properties can be flexibly tuned by adjusting the model parameters.
\end{abstract}

\maketitle

%

\section{Introduction}
Disordered systems have fascinating properties that can differ significantly from those of pure systems and can produce novel experimentally relevant effects.
For example, disorder frequently generates new universality classes, and can even completely alter the nature of phase transitions.  Most studies of disordered systems concern independent random disorder, but
if disorder correlations are present the issue arises of when and how these are relevant.
This paper focuses on spatial power-law-correlated quenched disorder that can arise, e.g., from linear or planar dislocations in a crystal, or for fluids in porous media. Such disorder is characterized by an exponent $a$, i.e., the correlation function of the defects decays as $g(r) \sim 1/r^a$.

A few guiding theories are available.
The Harris criterion \cite{Harris:criteria} states that weak uncorrelated disorder is irrelevant if the heat capacity exponent of the corresponding pure, disorder-free system is negative, $\alpha_{\rm{pure}}<0$ or the correlation length exponent $\nu_{\rm{pure}} > 2/d$ assuming the hyperscaling relation $d\nu=2-\alpha$. For power-law-correlated disorder, this is generalized to the Weinrib-Halperin (WH) criterion \cite{WH:criteria}, which predicts that weak disorder is irrelevant if the pure system satisfies $\nu_{\rm{pure}}>\max(2/a,2/d)$. Moreover, when disorder is relevant, a new disordered universality class is obtained if $a<d$ 
with exponents given by $\nu_d = 2/a$ and $\eta_d=0$. 
A number of papers have tested this criterion by numerical simulations.  While it seems reasonably well established that a new universality class is indeed obtained for the three-dimensional Ising model with oriented line disorder corresponding to $a=2$ \cite{parisi:Ising,russia:Ising,ukraine:Ising}, it is quite controversial whether $\nu_d=2/a$ holds. For example, it has been argued that the result is merely a first-order estimation, ignoring higher-order corrections \cite{PF1,PPF2,PPF3}. 
See Ref.\ \cite{ukraine:Ising} and the references therein for a more detailed discussion.

In this work we 
study power-law-correlated quenched disorder 
generated from equilibrium spin configurations of a pure, i.e., disorder-free 
zero-field Ising model at the phase transition.
This correlated disorder distribution avoids possible ambiguities due to, e.g., 
linear correlated defects crossing, and is much more straightforward to generate in simulations.
In addition, the decay exponent is $a\approx 1$ (see below) which is much smaller than that of line defects. Therefore, the predicted exponent $\nu_d \approx 2$ is extraordinarily large which is ideal for testing the WH criterion. 
A similar idea of using an auxiliary model to grow power-law-correlated disorder was used in previous studies of correlated random models. A three-dimensional (3D) Ising model with correlated random dilution was studied in Ref.~\cite{PhysRevE.62.191}. Potts models in 2D were studied in 
Refs.~\cite{Chatelain_2013,Chris:GP}. 
Here we study the different case of a random bond Ising model in 3D.

To define the disorder distribution, equilibrium 
spin configurations of a pure Ising model at the critical point are mapped onto
quenched random couplings of a disordered Ising system that by construction become power-law correlated.
We discuss the generating method in detail in the following sections. 
The random couplings are defined to take two values corresponding to the two 
spin orientations in the underlying Ising configuration.
While this disorder model does not fulfill the WH assumption of a Gaussian disorder distribution, 
the WH results are still useful guides and it is of interest to compare results. 
For simplicity we restrict to ferromagnetic couplings in this paper, i.e., 
there is no frustration and the ground state is the same 
ferromagnetic state as for the pure system.
The power-law decay of the random couplings to leading order 
follows from the pure spin correlation function exponent $a=d-2+\eta_{\rm{pure}}$. 
For the pure Ising model in three dimensions $a=1.036298(2)$, and $a<d$ is satisfied. 
The WH criterion therefore predicts for the disordered system a single phase transition with a new exponent $\nu_d\approx1.9299$. 
The critical cluster distribution leads to formation of power-law-correlated 
domains of strong and weak bonds in the random Ising model.
Here this model is studied theoretically to demonstrate what effects are in principle possible, 
without considering how it may be realized in practice.


The main purpose of the paper is to investigate the universality class of the phase transition in the three-dimensional Ising model with correlated bond disorder generated from pure Ising configurations, and in particular to identify the role of disorder correlations by comparing several different but related models. We use Monte Carlo simulations and finite-size scaling to study critical properties at the transition.
For correlated disorder a new universality class of the Ising phase transition emerges as expected from the WH theory,
but the values of the exponents do not follow the WH results.  
Notably the finite scaling approach to the thermodynamic limit is unusual
in the following sense. Quantities such as the susceptibility and heat capacity usually have a single rounded and finite peak close to the transition temperature for finite system sizes. 
Here instead such disorder-averaged quantities can obtain peaks at two different temperatures for finite systems. The two peak temperatures are system size dependent and merge in the thermodynamic limit,
so the infinite system has a single phase transition as expected from WH theory.
In addition to the correlated disorder model, we also study a layer model with the two values of bonds spatially separated to layers. The layer model is not disordered and 
trivially has two thermodynamic transitions and double peaks in the susceptibility that are not finite-size effects.  By selecting the values of the couplings and the number of layers, great flexibility to engineer the 
thermodynamic properties of the system is demonstrated. In particular several susceptibility peaks can be 
produced which is potentially relevant for magnetic applications.

An additional purpose is to investigate the performance of two different Monte Carlo algorithms,
the parallel tempering (PT) method \cite{ptmc1,ptmc2,Hukushima:PT} 
and the recently introduced population annealing (PA) method
\cite{Hukushima:PA,Zhou:PA,Machta:PA,Wang:PA,Weigel:PA}.
These methods have been used extensively to simulate spin-glass problems and similar problems with frustration 
and complex ground states where it has been found that PA and PT have similar efficiency \cite{Wang:GS,Wang:PA}.
Here we consider the case of correlated disorder with no frustration that preserves the ferromagnetic ground state and compare the performance of both methods.

The paper is organized as follows. We first discuss the models, observables, and simulation methods in Sec.~\ref{mm}, followed by numerical results in Sec.~\ref{results}. Concluding remarks are stated in Sec.~\ref{cc}.

\section{Models, observables and methods}
\label{mm}

\subsection{Models}

The Ising Hamiltonian is
\begin{eqnarray}
H = -\sum_{\langle ij \rangle} J_{ij} S_i S_j,
\end{eqnarray}
where $S_i = \pm 1$ are Ising spins and the summation is over nearest neighbours on a three-dimensional cubic lattice with side length $L$ and $N=L^3$ sites.
For the pure system with $J_{ij}=1$ the
critical temperature as well as critical exponents are known to high precision.
We take $\beta_{\rm c, pure}=0.22165455(3)$ obtained from Monte Carlo simulations \cite{deng:Ising} 
to generate critical configurations of the pure system. 
The conformal bootstrap method provides high precision estimates of critical exponents for the Ising model
given by $\nu_{\rm pure}=0.629971(4)$ and $\eta_{\rm pure}=0.036298(2)$ \cite{kos:exponents}. 
We refer to these quantities for the disordered system as $\beta_c$, $\nu_d$ and $\eta_d$, respectively.

The correlated quenched random couplings $J_{ij}$ are defined as follows:
\begin{enumerate}
\item Simulate the pure 3D Ising model and generate equilibrium configurations at the phase transition.
\item Define a set of random coupling constants from an equilibrium spin configuration of the pure model.
For bonds within spin-up clusters set $J_{ij}=2$, and 
otherwise $J_{ij}=1$, i.e., either within spin-down clusters and at cluster boundaries.
%
More precisely, $J_{ij}=1+(S_i+1)(S_j+1)/4$.
\end{enumerate}
The resulting values $J_1=1, J_2=2$ are fixed unless otherwise specified.
We refer to $J_1$ as weak bonds and $J_2$ as strong bonds. 
This method to define random couplings from underlying pure Ising configurations is not unique. The method leads to an asymmetry in the fraction of $J_1$ and $J_2$ bonds due to the treatment of the interfaces. It is possible to eliminate this asymmetry. One simple way is to let a single spin determine the bonds in the forward directions of each axis. For example, set a forward bond to $J_2$ if a spin takes the value $1$, and $J_1$ if it 
is $-1$. We call the asymmetric version pair disorder, and the forward version forward disorder. We simulated these and some other disorder models with variations of the short-range details and find they have similar properties. The data shown here are for pair disorder unless otherwise specified.


We considered a few variations in the definition of the bond disorder distribution.
If the disorder-generating Ising configurations have no restriction in the net magnetization,
the different disorder realizations will obtain varying numbers of strong and weak bonds,
which we call the unrestricted or $M\ne 0$ disorder.
If the generating spins are restricted to zero magnetization,
each disorder realization has equal numbers of strong and weak bonds.
We call this restricted or $M=0$ disorder.



For comparison with the random distribution of the couplings, 
we also considered a layer model which is not disordered by distributing 
couplings in a non-random way.
Here all couplings are assigned to $J_1$ in the upper half of the system, and 
$J_2$ in the lower half.  This is essentially two pure Ising models connected by flat interfaces between
regions with weak and strong bonds.  The layer model can be generalized in different ways, e.g., by changing sizes of layers, adding more layers and using more values of the couplings.  
If the number of spins in each layer is proportional to the total volume $L^3$, the ratio of the number of interface to bulk couplings disappears as $1/L$, and each layer trivially obtains its own thermodynamic bulk Ising transition at a critical temperature related to the coupling constant in the layer.  
Thus the $n$-layer model can have a sequence of $n$ transitions where the net magnetization changes 
in a staircase manner as temperature is varied.



\subsection{Observables and methods}

The main observables are the absolute value of the magnetization density $m$, 
Binder ratio $g$, magnetic susceptibility $\chi$, and heat capacity $c$.
These quantities are defined as:
\begin{eqnarray}
m &=& \frac{1}{N} \left| \sum_i S_i \right |, \\
g &=& \left[ \frac{\langle m^4 \rangle}{\langle m^2 \rangle^2} \right], \\
\chi &=& \beta N [\langle m^2 \rangle -\langle |m| \rangle^2 ], \\
c &=& \frac{\beta^2}{N} [\langle H^2 \rangle - \langle H \rangle^2 ].
\end{eqnarray}
Averages are performed over thermal fluctuations denoted $\langle ... \rangle$ and over quenched disorder denoted $[ ... ]$.

\begin{table*}
\caption{
Simulation parameters of the three-dimensional Ising model using population annealing (PA) and parallel tempering (PT) for the pure and disordered system with pair disorder. $L$ is the linear system size, $R$ is the number of replicas, $N_T$ is the number of temperatures, and $N_S$ is the number of sweeps. The temperature range is 
$\beta \in [\beta_{\rm{min}}, \beta_{\rm{max}}]$ and the Wolff updates are applied in the interval $\beta \in [\beta_1, \beta_2]$. Finally, $M$ is the number of independent runs or disorder realizations studied.
\label{para}
}
\begin{tabular*}{\textwidth}{@{\extracolsep{\fill}} l c c c c c c c c c r}
\hline
\hline
$\rm{System}$ &Algorithm &$L$  &$R$ &$N_T$ &$N_S$ &$\beta_{\rm{min}}$ &$\beta_{\rm{max}}$ &$\beta_1$ &$\beta_2$ &$M$ \\
\hline
pure &PA &$\{4, 6, 8\}$  &$8\times10^5$ &$101$ &$10$ &$0$ &$0.4$ &$0.2$ &$0.3$ &$20$ \\
pure &PA &$\{10, 12, 14, 16, 18\}$ &$1.6\times10^6$ &$201$ &$10$ &$0$ &$0.4$ &$0.2$ &$0.3$ &$20$ \\
pure &PA &$\{20, 24\}$ &$1.6\times10^6$ &$301$ &$10$ &$0$ &$0.4$ &$0.2$ &$0.3$ &$20$ \\
pure &PA &$\{28, 32\}$ &$1.6\times10^6$ &$301$ &$10$ &$0$ &$0.25$ &$0.2$ &$0.25$ &$20$ \\
pure &PT &$\{36, 40, 44, 50\}$ &$16$ &$-$ &$6\times10^6$ &$0.2$ &$0.24$ &$0.2$ &$0.24$ &$20$ \\
disorder &PA &$\{4, 6, 8\}$  &$8\times10^5$ &$101$ &$10$ &$0$ &$0.4$ &$0.1$ &$0.3$ &$5000$ \\
disorder &PA &$\{10\}$ &$1.2\times10^6$ &$201$ &$10$ &$0$ &$0.4$ &$0.1$ &$0.3$ &$3000$ \\
disorder &PA &$\{12, 14, 16\}$ &$1.2\times10^6$ &$201$ &$10$ &$0$ &$0.3$ &$0.1$ &$0.2$ &$3000$ \\
disorder &PA &$\{18, 20\}$ &$1.2\times10^6$ &$201$ &$10$ &$0$ &$0.22$ &$0.1$ &$0.2$ &$2000$ \\
disorder &PT &$\{24, 28, 32\}$ &$32$ &$-$ &$1.1\times10^6$ &$0.13$ &$0.15$ &$0.13$ &$0.15$ &$2000$ \\
disorder &PT &$\{36, 40, 44, 50\}$ &$32$ &$-$ &$2.2\times10^6$ &$0.13$ &$0.15$ &$0.13$ &$0.15$ &$2000$ \\
\hline
\hline
\end{tabular*}
\end{table*}

The finite-size scaling relations for computing critical exponents are summarized as follows:
\begin{eqnarray}
\label{gscaling}
g(t) &=& g(tL^{1/\nu}), \\
\label{chiscaling}
\chi(t) &=& L^{2-\eta}\chi(tL^{1/\nu}), \\
\label{gTscaling0}
g_T(t=0) &=& \left. \frac{\partial g}{\partial T} \right|_{t=0} \sim L^{1/\nu}, \\
\label{chiscaling0}
\chi(t=0) &\sim& L^{2-\eta},
\label{scaling}
\end{eqnarray}
where $t=(T-T_c)/T_c$ is the reduced temperature and $T_c$ is the phase transition temperature.
We measure $g_T$ using its direct estimator 
$\partial \mean{O}/\partial T = \beta^2 \left[ \mean{OH}-\mean{O}\mean{H} \right]$ 
where $O$ is any temperature independent quantity, here $m^2$ or $m^4$.
%
%

Our simulations are carried out using two different Monte Carlo (MC) methods,
population annealing \cite{Hukushima:PA,Zhou:PA,Machta:PA,Wang:PA,Weigel:PA} 
and parallel tempering \cite{ptmc1,ptmc2,Hukushima:PT} with hybrid Metropolis and Wolff updates. 
The two methods give consistent results, and both are in agreement with the known results of the pure system
for the transition temperature and the critical exponents. Since PT is widely used and well known, here we only discuss the relatively new PA method.

The PA method works as follows.
We initialize $R$ random configurations or replicas at $\beta=0$. The population of replicas are cooled gradually following an annealing schedule. When the temperature is decreased from $\beta$ to $\beta'$, a replica is copied with the expectation number $n_i=\exp[-(\beta'-\beta)E_i]/Q$, where $E_i$ is the energy of replica $i$ and $Q=(1/R)\sum_i \exp[-(\beta'-\beta)E_i]$ is a normalization factor to maintain the population size approximately the same throughout the annealing. In our simulation, the number of copies is either the floor or the ceiling of $n_i$ to minimize fluctuations with the proper probabilities to give the correct mean value. After this resampling step, $N_S$ sweeps using the Metropolis algorithm is applied to each replica. In the temperature range where the Wolff update \cite{Wolff} is efficient, we also apply $N_S$ Wolff updates to each replica. The two hybrid updates alternate with one Metropolis sweep and one Wolff update. The annealing process continues with resampling and Monte Carlo sweeps until reaching the lowest temperature.

Our equilibration criterion for PA is based on the family entropy, which quantifies the diversity of the population. In the initial population, each replica is assigned a family name $1, 2, 3, ..., R$. The family name is copied along with a replica in the annealing process. At each stage of the simulation, we can collect the fraction of each family name in the population $\{f_i\}$ and the family entropy $S_f$ is then the regular Gibbs entropy $S_f = -\sum_i f_i \ln (f_i)$ \cite{Wang:TBC,Wang:PA}. The larger $S_f$, the better the equilibration. We require each annealing to satisfy $S_f \geq \ln(100)$. To prevent correlations when generating the disordered instances, we have only recorded $500$ configurations from the much larger population for each simulation of the pure system and more configurations are collected using independent runs. 
For PT we used at least $10^5$ sweeps for thermalization before data collection. We record one equilibrium configuration for every $10^4$ sweeps at each temperature and also save $500$ configurations for each independent run. The simulation parameters are summarized in Table.~\ref{para}. Note that PA is massively parallel as different replicas can be updated independently. The PA is implemented with distributed-memory MPI parallel computing \cite{comment:mpi} and PT is implemented with shared-memory OpenMP parallel computing \cite{comment:openmp}.

\section{Numerical Results}
\label{results}

In this section, we first demonstrate that there is a single phase transition in three dimensions. Next, we study the critical behavior of the transition and compare with the WH criterion. Finally, we discuss the layer model and compare the efficiency of the PA and PT methods.

\subsection{Correlated disorder}
\label{dm}

\begin{figure}[htb]
	\begin{center}
		\includegraphics[width=\columnwidth]{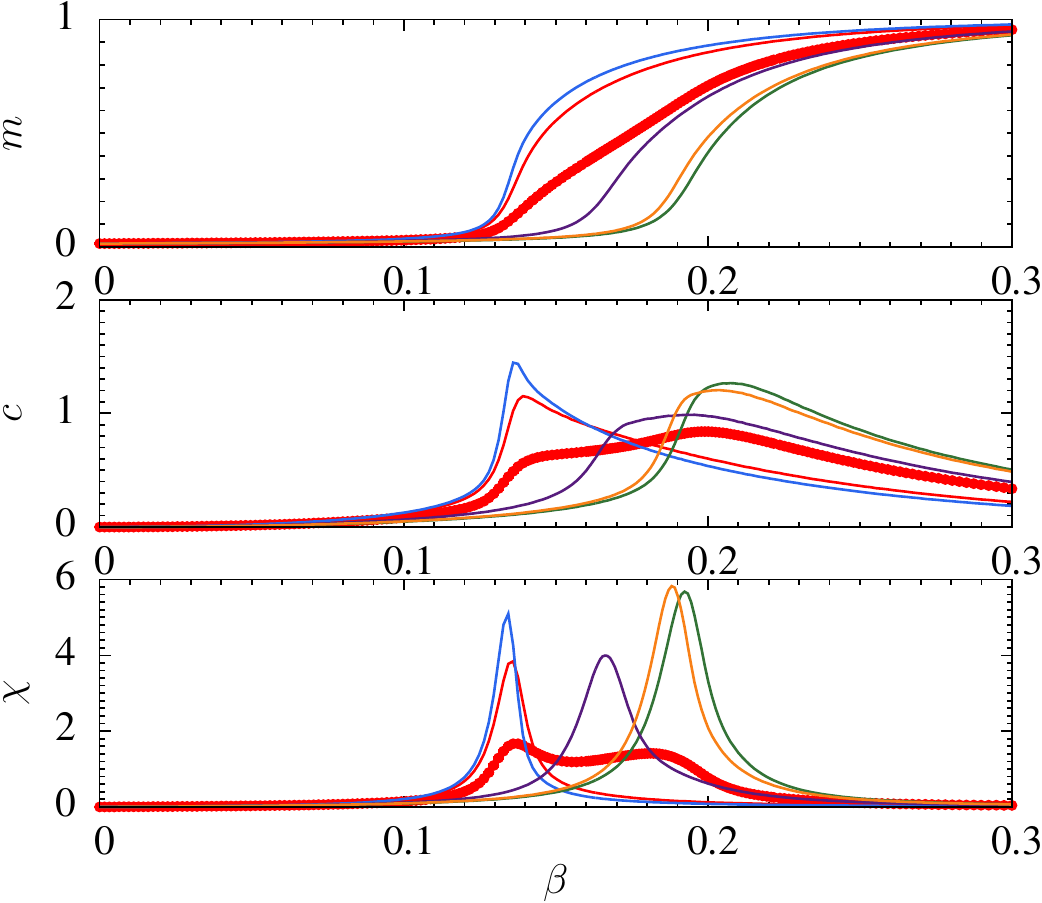}
		\put (-218,198) {(a)}
		\put (-218,132) {(b)}
		\put (-218,66) {(c)}
		\caption{
			Typical and disorder-averaged results for (a) absolute value of the magnetization, (b) heat capacity, and 
			(c) susceptibility of the correlated disorder model for system size $L=14$.
			The red dots are the disorder-averaged data and the thin curves are for five individual disorder realizations.
			The average magnetization has a finite-size rounded two step temperature dependence.  
			The 
			disorder averages in (b) and (c) have peaks at the two magnetization steps, 
			while individual disorder realizations have one peak. 
		}
		\label{TP}
	\end{center}
\end{figure}

We first show some typical and disorder-averaged MC results for the magnetization, heat capacity and susceptibility for the correlated disorder model 
for system size $L=14$ in Fig.~\ref{TP}.
The disorder averages are marked with red dots and form thick red curves. The thin curves are data for five individual disorder samples. 
The most noticeable feature of the disorder averaged curves is the two peak structure present both for the heat capacity and
susceptibility, indicating two characteristic temperatures. However, each individual sample has only one peak or transition.
The emergence of this double peak structure is one main result of this paper and will now be analyzed in more detail.


\begin{figure}[htb]
\begin{center}
\includegraphics[width=\columnwidth]{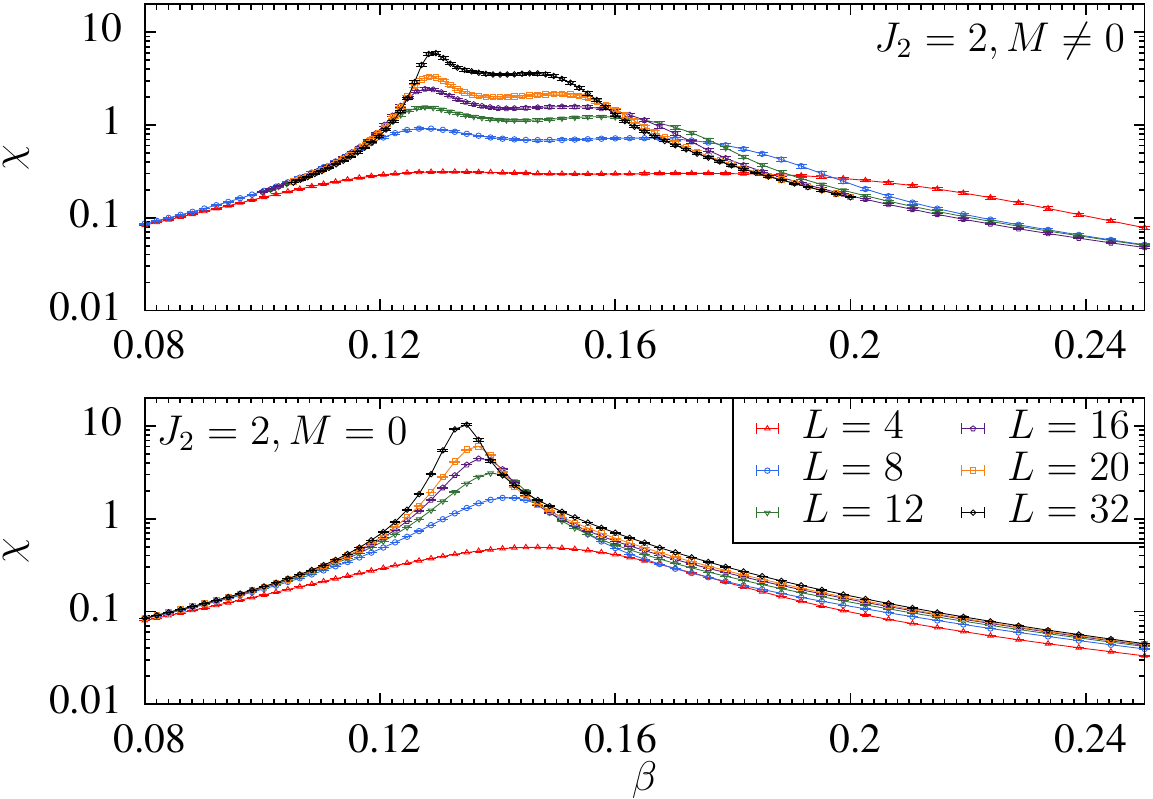}
\put (-211,160) {(a)}
\put (-211,26) {(b)} \\
\includegraphics[width=\columnwidth]{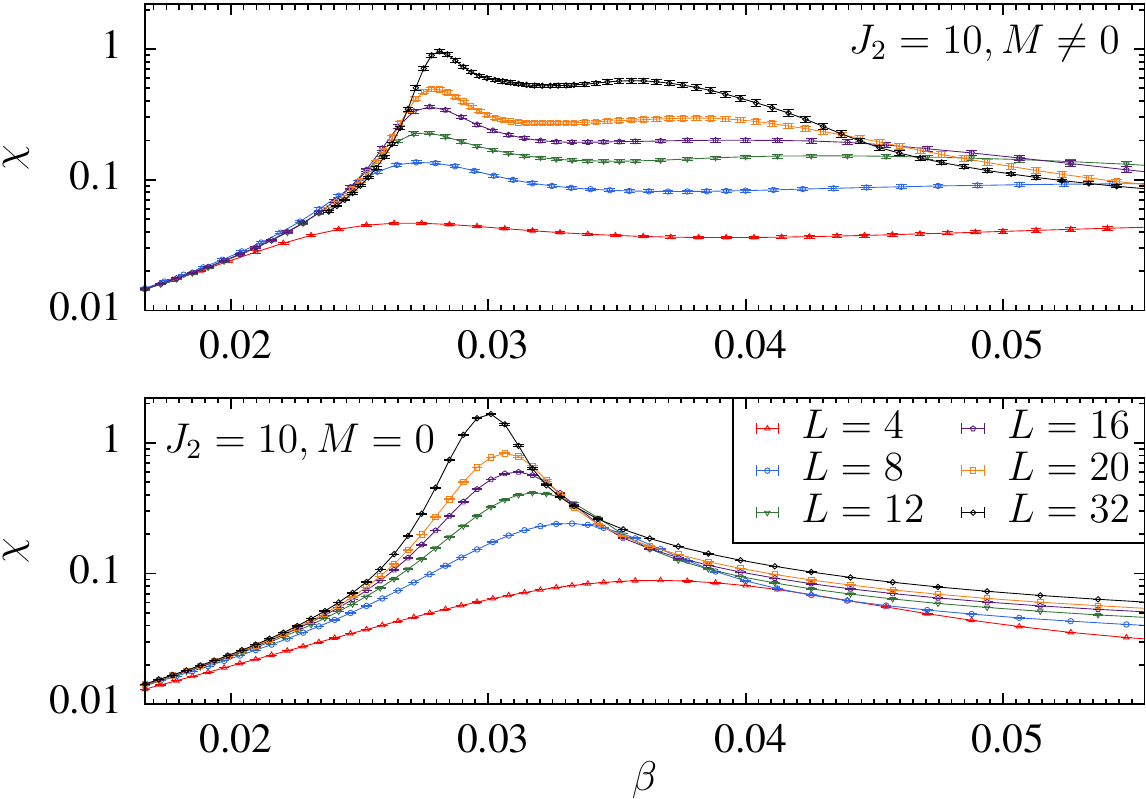}
\put (-211,160) {(c)}
\put (-211,34) {(d)}
\caption{
Disorder averaged susceptibility for comparing cluster disorder $(M\neq0$) and restricted cluster disorder ($M=0$). Here, we have examined two strong bond strengths and used the forward disorder. 
The figure illustrates that the two ensembles of disorder have the same thermodynamic limit, 
and there is therefore only one phase transition.
}
\label{chi2}
\end{center}
\end{figure}

Interestingly, the two peak structures after careful inspections are finite-size effects and there is a single phase transition in the thermodynamic limit $L\to\infty$.
This is firstly indicated from the fact that each individual disorder realization has only one peak.
The peak temperature reflects the average strength of the bonds of the disorder realization. 
Moreover, the two peak temperatures of the disorder averages move closer together as the system size increases; see Fig.~\ref{chi2} for more details. 
The peak at low (high) $\beta$ is associated with disorder realizations with
a majority of strong (weak) bonds.

To further demonstrate the two peaks are finite-size effects, we also studied a restricted disorder model which has always equally many strong and weak bonds in the disorder generating pure spin model. Just as the choice of thermodynamic ensemble does not change the properties of phase transitions, 
changing the disorder generating ensemble is not expected to 
affect the phase transition of the disordered system in the thermodynamic limit.   
This was investigated and confirmed in Ref.~\cite{PhysRevE.65.057104}.
Hence we expect the critical properties and the transition temperature to be the same for both disorder distributions,
although the finite-size properties are substantially different.
Results for the disorder-averaged susceptibility of the restricted disorder model are shown in Fig.~\ref{chi2} for the forward models with both $J_2=2$ and $J_2=10$.
Clearly only one peak is obtained for any system size and the same is true for the individual disorders. It is remarkable that this is the case even for the extraordinarily large strong bonds $J_2=10$.
From the figure it is also plausible that the transition temperatures of the two different 
disorder generating ensembles converge to the same values for $L\to\infty$. Here, the unrestricted ensemble has the two peaks originated from the two-peak distribution of magnetization of the underlying pure model, and in the restricted ensemble this is eliminated. This concludes the demonstration that the two peaks found above are finite-size effects caused by the disorder averaging procedure and the 
disorder generation method, and in the thermodynamic limit there is a single transition.

\subsection{Critical exponents}




The next step is to study the finite-size scaling of the phase transition. The main objective is to estimate the critical exponents at the transition by finite-size scaling of the susceptibility and Binder parameter derivative.
We consider finite-size scaling at the thermodynamic critical temperature. 
According to Eqs.~(\ref{gTscaling0}-\ref{chiscaling0}) the scaling at the critical temperature 
is $\chi\sim L^{2-\eta}$ and $g_T\sim L^{1/\nu}$. We have used the unrestricted pair disorder for this major large-scale simulation.

\begin{figure}[htb]
\begin{center}
\includegraphics[width=\columnwidth]{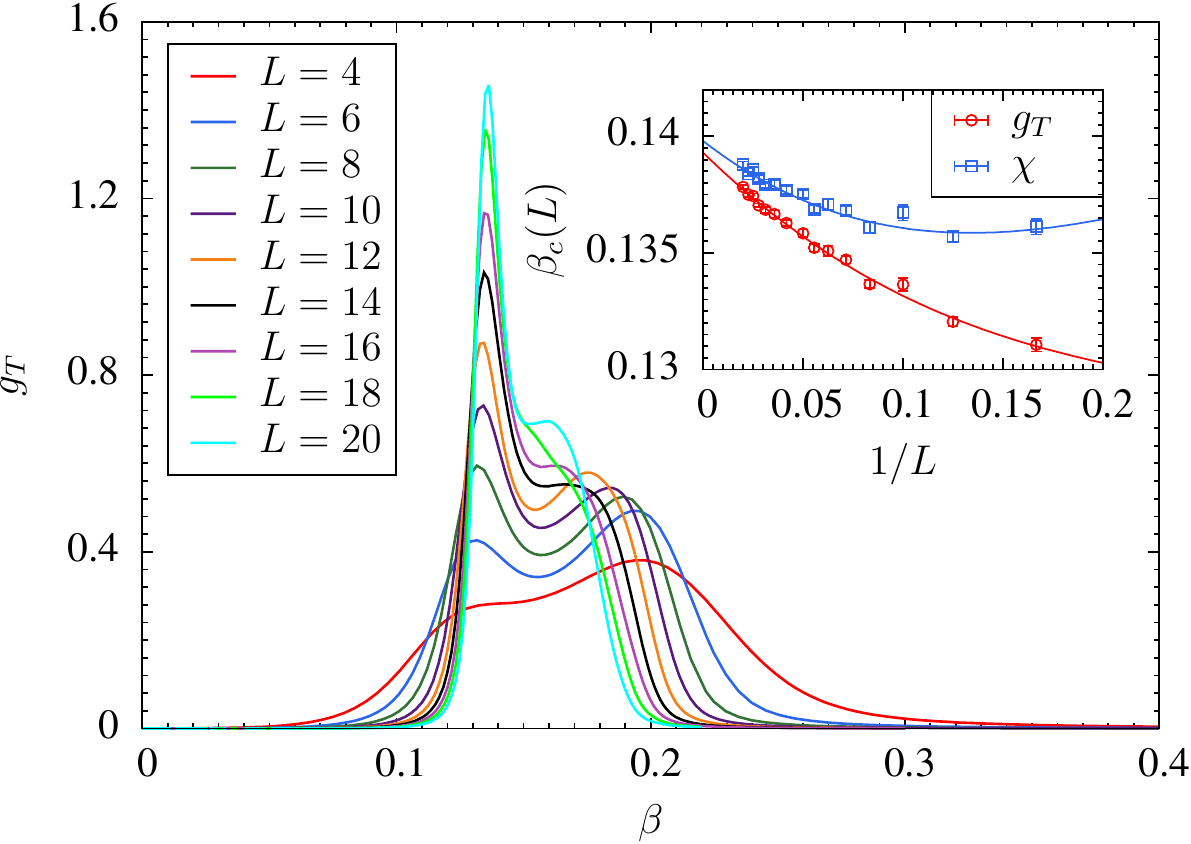}
\caption{
Disorder averaged $g_T$ vs. $\beta$ has a two peak structure for small system sizes.
The left peak dominates for large system size and is used to estimate the size dependent transition temperature.
A similar estimate is done for $\chi$.
Inset: The extracted pseudo transition temperature (squares) of the left peak vs. $1/L$ along with a cubic fit
(curves), giving $\beta_c=0.1393(3)$ and $0.1398(4)$ from the fits of $g_T$ and $\chi$, respectively. The $L=4$ data is omitted from the fit as there is no clear left peak for this size.
}
\label{GTBC}
\end{center}
\end{figure}

The first step is to estimate the thermodynamic transition temperature.
We use the major high-temperature peaks of $g_T$ and $\chi$ to estimate $\beta_c$.
%
%
%
%
Data for $g_T$ vs. $\beta=1/T$ for a series of sizes are shown in the main panel of Fig.~\ref{GTBC}. 
Similar behavior is found for $\chi$ (data not shown). 
It is seen that two peaks obtained for small sizes successively merge upon increasing the system size,
and the left peak at low $\beta$ dominates for large system sizes.
Hence the temperature of left peak should extrapolate to the thermodynamic transition temperature 
for $L\to\infty$.
Using cubic spline interpolation a pseudo transition inverse temperature $\beta_c(L)$ is 
computed by locating the maximum for each quantity and system size. 
Errorbars are estimated using the bootstrap method.  
Ideally one may would like to estimate the thermodynamic transition temperature using the scaling relation $\beta_c(L) = \beta_c + a/L^{1/\nu_d}$, but this requires simultaneous fits of $\beta_c$ and $\nu_d$ which leads to large statistical errors for this disordered model. 
We instead adopt a cubic polynomial fit in $1/L$ and extrapolate to $1/L=0$; see the inset in Fig.~\ref{GTBC}.
The validity of this method was verified using the scaling fits assuming $\nu_d=1.13$ estimated below, with consistent results within errorbars.
We also verified that this method reproduces known results reliably for 
both the critical temperature and critical exponents of the corresponding pure model.
The results for the correlated disordered model are $\beta_c=0.1393(3)$ and $0.1398(4)$ 
from the fits of $g_T$ and $\chi$, respectively. 
The results agree within error bars and we combine the two and estimate $\beta_c=0.1396(3)$, 
which is close to the simple estimate 
$(2/3)\beta_{\rm c,pure}=0.1478$ given by the mean bond value.

%


\begin{figure}[htb]
\begin{center}
\includegraphics[width=\columnwidth]{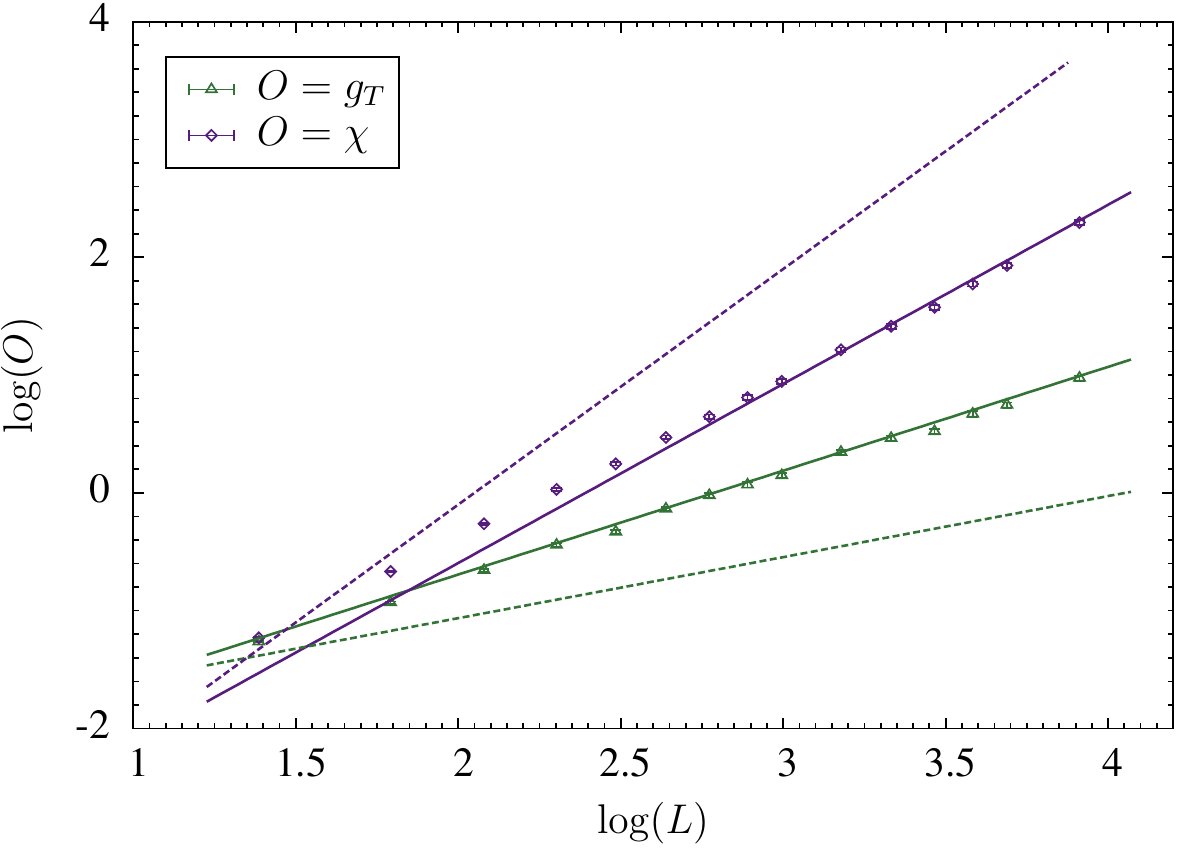}
\put (-14,30) {(a)} \\
\includegraphics[width=\columnwidth]{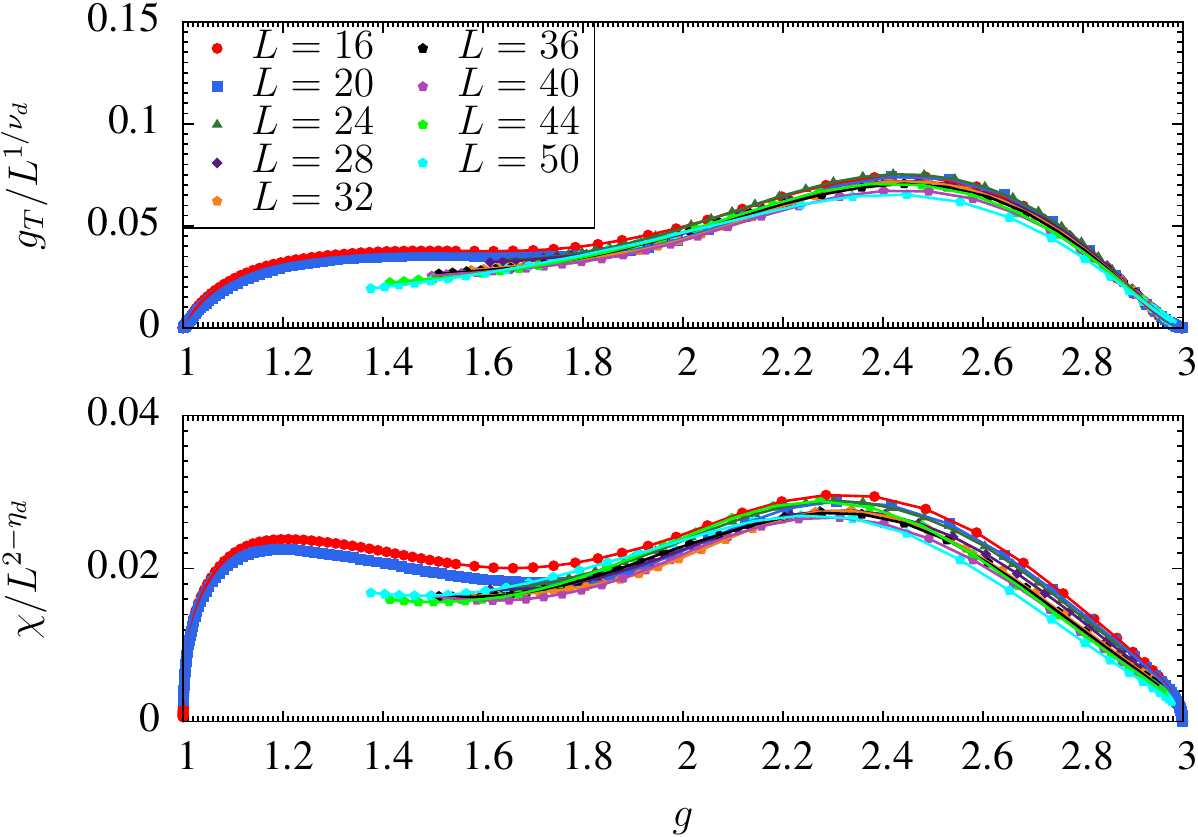}
\put (-16,158) {(b)}
\put (-16,77) {(c)}
\caption{
(a) Scaling of $g_T$ and $\chi$ with system size $L$ at the thermodynamic transition temperature [(a)]. The linear solid fits are done using the seven largest sizes, and the dashed lines are guiding lines with the predicted WH slopes. 
(b-c) 
finite-size scaling data collapse of scaled quantities vs. Binder ratio $g$.  
The sizes included are $L=16, 20, 24, 28, 32, 36, 40, 44, 50$. 
}
\label{EP}
\end{center}
\end{figure}

Our main results for critical exponents are summarized in Fig.~\ref{EP}. 
Notice that to a good approximation a power law is obtained for $g_T$, while 
substantial deviations from a pure power law are found in the data curve for 
$\chi$ for small system sizes, indicating the presence of scaling corrections.
%
Including data points for the seven largest system sizes where a power-law fit is 
reasonably justified gives 
\begin{eqnarray}
\nu_d &=& 1.13(5), \nonumber \\
\eta_d &=& 0.48(3).
\label{exponents}
\end{eqnarray}
Error bars are bootstrap estimates.
Note that the exponents differ significantly from both the 
pure universality class and with the WH results.
In particular, the exponent $\nu_d$ sits in between these two values. 
This behavior is similar to recent results for models with line 
defects~\cite{russia:Ising,ukraine:Ising}.

For uncorrelated disorder recent simulation estimates~\cite{Theodorakis2011} of the exponents for the 3D 
Ising model with bond disorder are
$\nu=0.685(7)$ and $\gamma/\nu=1.964(9)$, which gives 
$\eta=2-\gamma/\nu=0.036(9)$. 
These values are significantly different from the values obtained here for correlated disorder,
showing that the long-range correlations are relevant and change the universality class
as expected.





Next we consider data collapses to verify that the estimated exponents hold.
It is useful to plot $g_T$ and $\chi$ vs. $g$ in order to eliminate explicit temperature scaling.
The results are shown in Fig.~\ref{EP}. A scaling collapse onto a common function for each quantity is found around the phase transition. This confirms that the data fulfills the finite-size scaling assumption.


Summarizing, the correlated disorder model produces a new universality class that differs from the 
pure model. This agrees with the expectation from WH theory, but the critical exponents 
differ significantly from the WH prediction.

\subsection{Bilayer model}
\label{bm}

In the previous section, the 3D Ising model with critical correlated disorder has two peaks in the disorder-averaged susceptibility for finite system size that merge into one 
transition in the thermodynamic limit,
while individual disorder realizations have a single peak.
This motivates the problem of 
constructing a simpler disorder-free model with similar double peak properties.
Classifying spins to two layers depending on the coordinate $z$-axis
such that each layer occupies half the system sites.
In the top layer the bonds of each spin in the forward direction are fixed to $J_{\rm{up}} = 1$, and in the bottom layer the bonds similarly are all set to, e.g., $J_{\rm{dn}}=2$. 
In contrast to the random model studied in the previous section there is no disorder.
Periodic boundary conditions are used in all directions, so
there are two boundaries between strong and weak coupled spins.

\begin{figure}[htb]
\begin{center}
\includegraphics[width=\columnwidth]{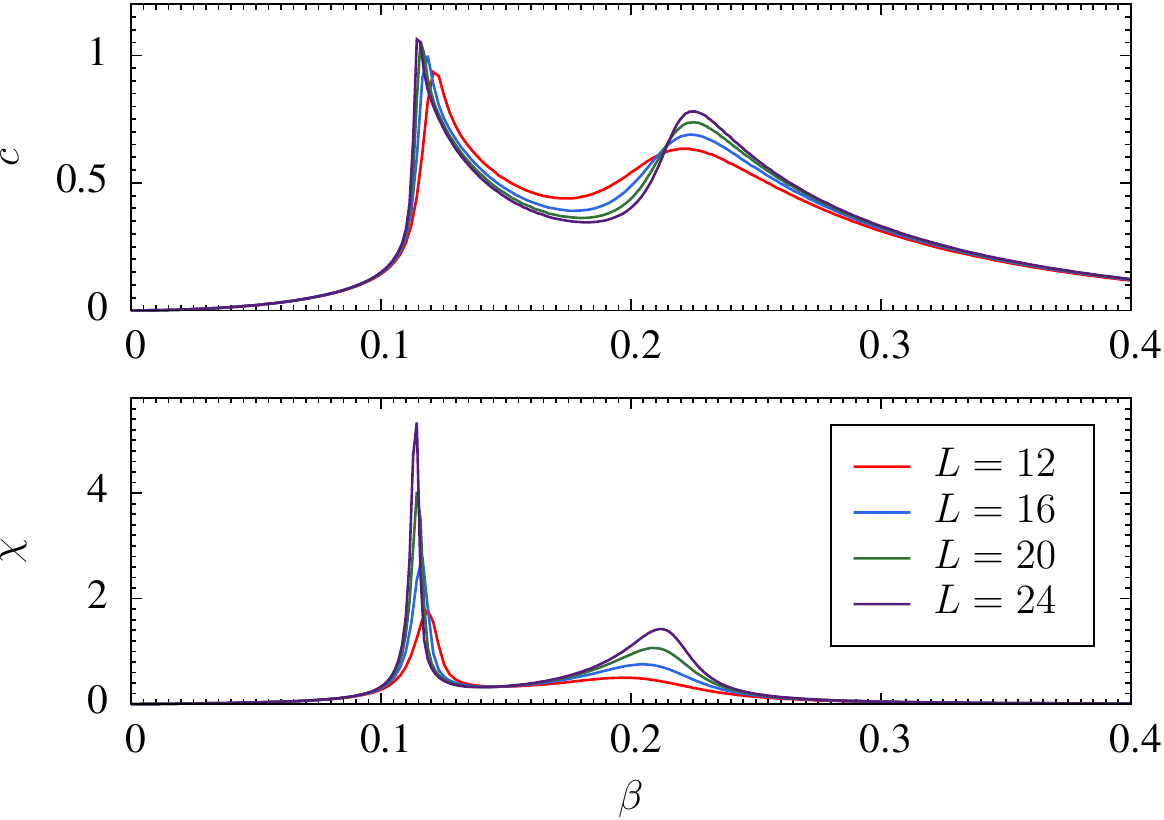}
\put (-214,162) {(a)}
\put (-214,78) {(b)}
\caption{Monte Carlo data for the bilayer Ising model with $J_{\rm{up}}=1$ and $J_{\rm{down}}=2$.
(a) Heat capacity. 
(b) Susceptibility.
In contrast to the cluster disorder model the 
two peaks of the layer model indicate two phase transitions.
Errorbars are small and omitted for clarity.
}
\label{BL}
\end{center}
\end{figure}

We studied a series of system sizes for $J_{\rm{dn}}=2$, and MC data for the heat capacity and susceptibility are shown in Fig.~\ref{BL} for four representative sizes. 
There are clearly two peaks emerging, one at $\beta_{c,\rm{pure}}$ and the other at $\beta_{c,\rm{pure}}/2$.
The layers are only coupled at the interfaces
so that 
each layer trivially has an independent phase transition in the thermodynamic limit. 
Thus the two peak structure in the correlated disorder model can be reproduced in the simple layer model,
but in contrast to the disordered case, the peaks of the layer model do not merge in the thermodynamic limit.

%
%

\begin{figure}[htb]
\begin{center}
\includegraphics[width=\columnwidth]{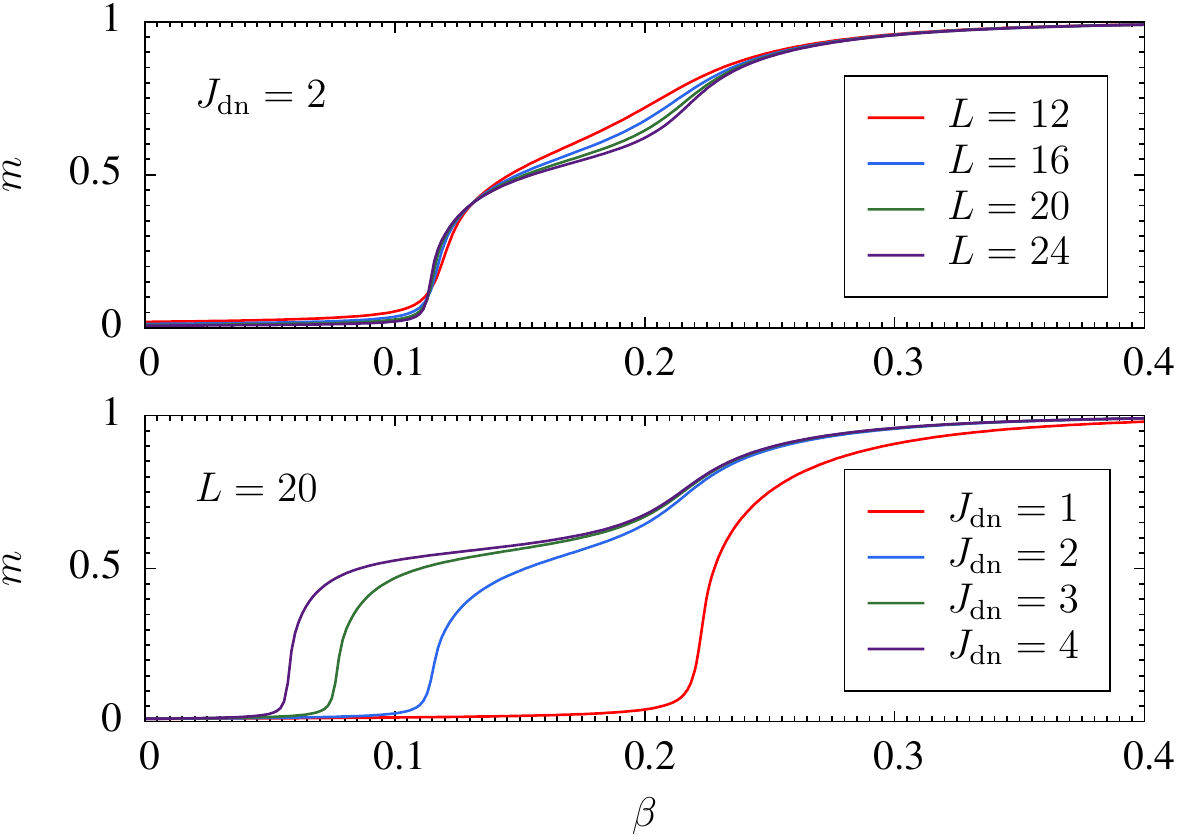}
\put (-214,111) {(a)}
\put (-214,27) {(b)}
\caption{
(a) Magnetization curve for different sizes for $J_{\rm{up}}=1$ and $J_{\rm{dn}}=2$. (b) Dependence of the magnetization curve on $J_{\rm{dn}}$ for fixed $J_{\rm{up}}=1$ and system size $L=20$.
%
Errorbars are small and omitted for clarity.
}
\label{BL2}
\end{center}
\end{figure}

A useful feature of the layer model is that 
%
%
%
with two phase transitions the shape of the magnetization curve can be designed by selecting the coupling strengths. This is illustrated in Fig.~\ref{BL2} where the magnetization curve is shifted by tuning $J_{\rm{dn}}=1, 2, 3, 4$ while keeping $J_{\rm{up}}=1$ fixed. 
Since the system has two independent phase transitions the magnetization 
in the thermodynamic limit is given by
\begin{eqnarray}
m(\beta) = 1/2 [ m_0(\beta) + m_0(\beta J_{\rm{dn}}/J_{\rm{up}}) ],
\end{eqnarray}
where $m_0$ is the magnetization of the pure system.

It is straightforward to generalize to multiple layers, different layer volumes, etc.
This permits engineering the magnetization curve to a desired shape. 
For $p$ layers of strength $J_i$ and weights $\omega_i$, which is the fraction of the number of spins of layer $i$, the magnetization in the thermodynamic limit is generalized to
\begin{eqnarray}
m(\beta) = \sum_{i=1}^p \omega_i m_0(J_i\beta).
\end{eqnarray}

\subsection{Efficiency of PA and PT}

Our work has also changed our view regarding the efficiency of the PA and PT algorithms. The PA algorithm has been used successfully in a number of studies of spin glasses, a system that is both disordered and frustrated. It has been found the two algorithms are quite similar in efficiency in both equilibrium samplings \cite{Wang:PA} and optimizations \cite{Wang:GS}.  Here and in our recent work on two-dimensional monodisperse particles \cite{Wang:VC}, where frustration is either fully absent or much weaker, we find that PT has an advantage over PA for studying phase transitions.

The key difference is the much shorter equilibration and autocorrelation times for less disordered and frustrated systems, where one can quickly equilibrate and start data collection using PT near the phase transitions. On the other hand, PA has to do the ``temperature journey'' and a large portion of work is ``wasted'' in this process. 
For example, to simulate a pure model only at the critical temperature 
it is clearly less efficient to use PA. 
On the other hand for collecting data over a wide range of temperatures the two algorithms are again similar in efficiency. Therefore PT can have an advantage over PA for studying phase transitions of pure or weakly disordered and frustrated systems, but the efficiency is similar for studying a wide range of temperatures or complex energy landscapes. In addition, PA has certain interesting features not shared by PT such as being massively parallel.

\section{Conclusions}
\label{cc}

In common model magnets the order parameter signals a transition from a magnetically ordered to disordered state as temperature is increased,
and thermodynamic quantities such as the magnetization curve have little structure except for the singularity at the transition.
For finite size samples, for example nanosize magnetic particles relevant in applications, 
divergent quantities become rounded with a single peak at the transition.
Introduction of uncorrelated quenched random disorder may alter the universality class of the phase 
transition but the qualitative features remain similar to the pure case.  
In this paper we point out that presence of spatially correlated quenched disorder can alter this picture.

We consider power-law-correlated critical 
cluster disorder generated from the equilibrium states at a second-order phase transition. 
We apply correlated critical cluster disorder as random bond couplings of a 3D Ising model.
We find unusual double peaked disorder-averaged susceptibility and 
other fluctuation quantities for finite system sizes, 
but one phase transition in the thermodynamic limit.
The appearance of the double peaks depend on the method of generating disorder
but not on the disorder correlations. 
In particular, if the disorder generating configurations are restricted to zero magnetization,
then disorder averages become single peaked.

From finite-size scaling of MC data we obtain $\nu_d=1.13(5), \eta_d=0.48(3)$. 
For power-law-correlated disorder with sufficiently slow decay as in our 
correlated disorder model, WH theory predicts a new long range fixed point.
This prediction agrees with our simulation results but the predicted exponents 
disagree with our 
exponents suggesting that the WH results need higher-order corrections
to apply to our bimodal disorder distribution.

We also studied a bilayer model which has two susceptibility peaks, similar to the correlated disorder 
results.
However, contrary to the critical disorder model the double peak structure appears at a single sample level and persists in the thermodynamic limit and there are therefore two separate phase transitions.  
Notably, the shape of the magnetization vs. temperature 
curve can be modified by changing the parameters of the system.  
By straightforward generalization to multiple layers, we propose a way to engineer the 
magnetization curve almost at will by tuning the coupling strength and weights of different layers.

For 2D Potts models with a similar disorder construction as used here, Chatelain \cite{Chatelain_2013,Chris:GP}
obtained interesting hyperscaling violations related to lack of self averaging, and Griffiths phases. While we consider a different model, it is notable that we do not observe these effects in our 3D results. Using the scaling relation $m(t=0) \sim L^{-\beta_d/\nu_d}$, we estimate $\beta_d/\nu_d=0.73(4)$ again using the seven largest sizes. The pertinent magnetic hyperscaling ($\gamma_d/\nu_d=d-2\beta_d/\nu_d$) is well satisfied after applying $\gamma_d/\nu_d=2-\eta_d$.

The critical cluster disorder model can be readily applied in other dimensions and to other $O(N)$ models, e.g., by adding the Ising correlated disorder to the XY and Heisenberg models. It is also interesting to study spin glasses with such correlated disorder. Research work along these lines are currently in progress and will be reported in future publications.


\acknowledgments 
W.W.~acknowledges support from the Swedish Research Council Grant No.~642-2013-7837 and the
Goran Gustafsson Foundation for Research in Natural Sciences and Medicine. 
M.W.~acknowledges support from the Swedish Research Council Grant No.~621-2012-3984.
The computations were performed on resources
provided by the Swedish National Infrastructure for Computing (SNIC)
at the National Supercomputer Centre (NSC), the High Performance Computing Center North (HPC2N), 
and the PDC Center for High Performance Computing.

\bibliography{Refs}

\end{document}